\documentclass[11pt]{article}

\usepackage{amsmath,amssymb}
\usepackage{slashed}
\usepackage{xcolor} 
\usepackage{graphicx}
\usepackage{dcolumn} 
\usepackage{bm} 
\usepackage{epsfig}
\usepackage{epstopdf}
\usepackage{grffile}
\usepackage{color}
\usepackage{colordvi}
\usepackage{amsmath,amssymb}
\usepackage{rotating}
\usepackage{lscape}
\usepackage{cite}
\usepackage{float}
\usepackage{hyperref}

\newcommand{\head}[1]{\textnormal{\textbf{#1}}}

\def\Re{{\cal R \mskip-4mu \lower.1ex \hbox{\it e}\,}}
\def\Im{{\cal I \mskip-5mu \lower.1ex \hbox{\it m}\,}}

\def\tev{\,{\ifmmode\mathrm {TeV}\else TeV\fi}}
\def\gev{\,{\ifmmode\mathrm {GeV}\else GeV\fi}}
\def\mev{\,{\ifmmode\mathrm {MeV}\else MeV\fi}}
\def\to{\rightarrow}

\begin{document}

\begin{center}

\vspace*{15mm}
\vspace{1cm}
{\Large \bf Direct and Indirect Searches for Top-Higgs FCNC Couplings  }

\vspace{1cm}

{\bf  Hoda Hesari${}^a$, Hamzeh Khanpour${}^{b,a}$ and Mojtaba Mohammadi Najafabadi${}^a$ }

 \vspace*{0.5cm}

{\small\sl$^a\,$School of Particles and Accelerators, Institute for Research in Fundamental Sciences (IPM) P.O. Box 19395-5531, Tehran, Iran } \\
{\small\sl$^b\,$Department of Physics, University of Science and Technology of Mazandaran, P.O. Box 48518-78195, Behshahr, Iran } \\

\vspace*{.2cm}
\end{center}

\vspace*{10mm}

%
%
\begin{abstract}\label{abstract}
Large top quark flavor changing through neutral currents is expected by
many extensions of the standard model. Direct and indirect searches for flavor changing neutral currents (FCNC)
in the top quark decays to an up type quark (up,charm) and a Higgs
boson are presented.
We probe the observability of the top-Higgs FCNC couplings through the process
e$^-$e$^+ \rightarrow t (\rightarrow \ell \nu_{\ell} b) \,\, \bar{t} (\rightarrow q H)$, where $\ell$ = e, $\mu$ and $q$ reflects up and charm quarks.
It is shown that the branching ratio $Br(t\rightarrow qH)$ can be
probed down to $1.12\times 10^{-3}$ at $95\%$ C.L. at the center-of-mass energy of 500 GeV with
the integrated luminosity of 3000 fb$^{-1}$.
We also update the constraint on the top-Higgs FCNC coupling using
the electroweak precision observables related to $Z\rightarrow c\bar{c}$ decay.
\end{abstract}

\vspace*{3mm}

PACS Numbers:  13.66.-a, 14.65.Ha

{\bf Keywords}: Top quark, Higgs boson, Flavor changing neutral current.

\newpage

%
%
\section{Introduction}\label{Introduction}

The discovery of Higgs boson with a mass of about 125 GeV by the ATLAS and CMS experiments at the CERN-LHC~\cite{Aad:2012tfa,Chatrchyan:2012ufa},
has opened a window to search for new physics through precise measurements of the processes involving this particle.
In particular, precise measurements of Higgs boson couplings to the Standard Model (SM) particles and its mass provide
excellent opportunities for searches for the SM extensions.
The Higgs boson mass and couplings to fermions and gauge bosons have  been measured in various decay modes, and they
are found to be in agreement with the predictions of the SM within
uncertainties ~\cite{Khachatryan:2014jba,Khachatryan:2015ila,CMS:ril,CMS:yva,CMS:2015mda}.

Top quark, the heaviest element of the SM, has the largest Yukawa coupling to the Higgs boson.
With a mass of around 173.5 GeV,  comparable to the electroweak symmetry breaking scale, measurement of the top quark properties
would provide an appropriate probe for electroweak symmetry breaking mechanism.
Within the SM, Higgs boson couples to fermions via Yukawa interactions and thereby producing the mass terms.
There are no Flavor Changing Neutral Current (FCNC) transitions
mediated by the Higgs boson or
by the $Z$, $\gamma$, $g$ gauge bosons at tree level. In other words, no leading order transitions of
$t\rightarrow qH$ or $t\rightarrow qV$, where $q$ reflects up or charm
quarks and $V=\gamma,Z,g$, exists in the SM framework.
The SM contributions to the top quark FCNC occur at loop level with the expected branching ratios around 10$^{-15}$-10$^{-13}$~\cite{AguilarSaavedra:2004wm}.
Such FCNC transitions are highly suppressed due to the
Glashow-Iliopoulos-Maiani (GIM) mechanism~\cite{Glashow:1970gm} and
top quarks almost exclusively decay to a bottom quark and a $W$ boson~\cite{Khachatryan:2014nda,Eilam:1990zc,Mele:1998ag}.

However, in some SM extensions, suppression due to the GIM mechanism can be relaxed because of the
additional contributions of new particles in the loop diagrams and consequently larger branching ratios
of $t\rightarrow qH$ or $t\rightarrow qV$ are expected.
Quark singlet model \cite{AguilarSaavedra:2002ns,delAguila:1998tp}, two Higgs doublet models \cite{hd1,hd2,hd3,hd4,hd5,hd6},
the minimal supersymmetric standard model (MSSM) \cite{mssm1,mssm2,mssm3,
mssm4,mssm5,mssm6}, extra dimensions \cite{ed},
and natural composite Higgs models \cite{ch,gilad} are examples of the SM extensions in which significant enhancements of top quark FCNC appear.
Even, in type III of two Higgs doublet model without flavor conservation, the $t\rightarrow qH$ transitions appear at tree level. These extensions of
the SM can enhance the branching ratio of $t\rightarrow qH$ up to $10^{-5}$.
Consequently, measuring any excess in the branching ratios for top quark FCNC processes would be an indication to physics beyond the SM.
There are already very many studies on the probe of the FCNC processes
and anomalous couplings in the top quark sector in the literature
\cite{Wu:2014dba ,Khatibi:2014via,Agram:2013koa,AguilarSaavedra:2000db,AguilarSaavedra:2004wm,Larios:2004mx,Goldouzian:2014nha,Wang:2012gp,Durieux:2014xla,
Hesari:2014eua,Khanpour:2014xla,saav, Craig:2012vj,etesami,bernreuther,x1,x2,x3,x4,x5,x6,x7,x8,x9,x10,x11}.

Searches for the existence of physics beyond the SM can be performed either at high
energy colliders or using its indirect effects in higher
order processes.
In this paper, we perform direct and indirect probes for the top-Higgs FCNC couplings. We redo the calculations which have been performed in
 Ref.\cite{Larios:2004mx} on the effects of top-Higgs FCNC couplings
 in the electroweak precision observables of $Z$ boson and
update the upper limit on $Br(t\rightarrow cH)$ using the most recent measurements.

There are several proposals for a possible future e$^-$e$^+$ collider
\cite{Barklow:2015tja,Baer:2013cma,Brau:2012hv,Linssen:2012hp,Aicheler:2012bya,Gomez-Ceballos:2013zzn,CEPC-SPPC} which would provide precise
measurements in particular in the top quark sector and Higgs boson properties.
As a direct way to search for the top-Higgs FCNC interactions, we study the sensitivity of a future e$^-$e$^+$
collider via $t \bar{t}$ events at centre-of-mass energy of 500 GeV.
We consider the case that one of the top quarks decays to a $W$ boson and a bottom quark with leptonic decay of the W boson $(t \rightarrow \ell \nu_{\ell} b)$
and the other top quark decays anomalously $t \rightarrow q H$ ($q$ = $u$ and $c$).
We consider $H \rightarrow b \bar{b}$  decay mode as the Higgs boson decay into $b\bar{b}$ pairs has maximum
branching ratio ~\cite{Denner:2011mq} and high efficiency in tagging the jets originating from the
hadronization of bottom quarks can be achieved ~\cite{Baer:2013cma,Abramowicz:2013tzc,Suehara:2015ura}.
We provide the 95$\%$ C.L. upper limit on the branching ratio of $t \rightarrow q H$ for various b-quark tagging efficiencies.
There are several proposals for the center-of-mass energy and
the integrated luminosity for a future electron-positron collider in the literature~\cite{Asner:2013psa,Dawson:2013bba,Peskin:2013xra,Fedderke:2015txa}.
We give the results for the integrated luminosities of  300 and 3000 fb$^{-1}$ of data and the center-of-mass energy of $\sqrt{s}$ = 500 GeV.

This paper is organized as follows. In Section~\ref{Top-Higgs-FCNC-interactions},
we briefly describe the theoretical framework which we consider to study the top-Higgs FCNC interactions.
In Section~\ref{Recentstatus}, we review the current best limits on
top-Higgs FCNC processes from direct and indirect searches.
The update of the upper limit on $Br(t\rightarrow cH)$ using
electroweak precision observables of $Z$ boson is also presented in Section \ref{Recentstatus}.
In Section~\ref{Numerical-study}, we describe the Monte Carlo event
generation, detector simulation
for top pair production in electron-positron collisions
with FCNC decays of one of the top quarks ($t\rightarrow qH$). The 95$\%$ C.L. upper limits on the
branching ratio of $t\rightarrow q H$ at different integrated
luminosities and various b-tagging efficiencies are also presented in
this section.
Finally, our summary and conclusion are given in Section~\ref{Conclusions}.

%
%
\section{Theoretical framework}\label{Top-Higgs-FCNC-interactions}
The general effective Lagrangian describing the interaction of a light up type quark ($q$ = $u$, $c$) with the top quark and a Higgs boson can be written as
~\cite{AguilarSaavedra:2009mx}:
\begin{eqnarray}\label{fcnclagrangian}
\mathcal{L} = -\frac{g} {2\sqrt{2}}\sum_{q = u, c}g_{tqH} \bar{q}(g^{v}_{tqH} + g^{a}_{tqH}\gamma_{5})tH + h.c. \, ,
\end{eqnarray}
where the dimensionless real coefficient $g_{tqH}$ (with $q$ = $u$
and $c$) denotes the strength of the top-Higgs FCNC coupling, and $g$ is the
weak coupling constant.
The coefficients $g^{v}_{tqH}$ and $g^{a}_{tqH}$ are general complex
numbers with the normalization $|g^{v}_{tqH}|^{2} + |g^{a}_{tqH}|^{2}
= 1$.
Strong cancellations arising from the GIM mechanism cause a tiny value
for $g_{tqH}$ in the SM. In the SM framework, $g_{tqH}$ amounts to
$10^{-6}$ while in a big range of MSSM parameters space, a sizeable
value at the order of $10^{-2}$ is expected \cite{mssm4,saav}.

After neglecting the up and charm quark masses, the $t \rightarrow q
H$ and $t\rightarrow  b W$ widths at leading order can be written as:
\begin{eqnarray}
\Gamma (t \rightarrow q H)  =  \frac{\alpha}{32 \, s_W^2} |g_{tqH}|^2
\, m_t \left[ 1 - \frac{M_H^2}{m_t^2} \right]^2  \, , \nonumber
\Gamma(t \rightarrow b W) = \frac{\alpha |V_{tb}|^{2}}{16
  s_{W}^{2}}\frac{m_{t}^{3}}{m_{H}^{2}} \left(1-\frac{3m_{W}^{4}}{m_{t}^{4}}+\frac{2m_{W}^{6}}{m_{t}^{6}} \right),
\end{eqnarray}
where $\alpha$ is the fine structure constant, $V_{tb}$ is the CKM
matrix element, $s_W$ is the sine of the Weinberg angle,
$m_{t},m_{W}$ and $m_{H}$ are the top quark, $W$ boson, and Higgs boson masses, respectively.
We estimate the branching ratio of $t\rightarrow q H$ as the ratio of
$\Gamma (t \rightarrow q H)$ to the width of $t\rightarrow Wb$. It has the following form:
\begin{eqnarray}\label{branchingratio}
Br(t\rightarrow q H) = \frac{g_{tqH}^{2}}{2} \times
\frac{x^{2}}{1-3x^{4}+2x^{6}}\left(1-y^{2} \right)^{2} = 0.0274\times g_{tqH}^{2},
\end{eqnarray}
where $x = m_{W}/m_{t}$ and $y=m_{H}/m_{t}$.
For the calculations, we use  $m_{H}$=125.7 GeV, $m_{t}$=173.21 GeV, $\alpha=1/128$, and $m_{W}$= 80.38 GeV \cite{pdg}.

%
%
\section{Current constraints on $Br(t\rightarrow qH)$}\label{Recentstatus}

In this section, we review the currently available limits on the branching ratio of $t\rightarrow qH$
from the collider experiments as well as the indirect limits. We also
update the limits from observables related to $Z\rightarrow c\bar{c}$ decay.

{\it Direct limits:}
The ATLAS search for the $tqH$ FCNC is based on the top quark pair events with one top quark decays of $t \rightarrow q H$  ($H \rightarrow \gamma \gamma$)
and the standard decays of the other top quark. The analysis uses 4.7
fb$^{-1}$ and 20.3 fb$^{-1}$ integrated luminosity of
data collected at $\sqrt{s}$ = 7 and 8 TeV, respectively.
Assuming $m_{H}$ = 125.5 GeV, the observed limit on the branching ratio of $t \rightarrow q H$ at
 95$\%$ C.L. is  $7.9\times 10^{-3}$~\cite{atlastqh}. This analysis has set an upper limit of $5.1\times 10^{-3}$
at 95$\%$ C.L. on $Br(t\rightarrow cH)$.

The limits from the CMS experiment is based on an inclusive search involving lepton and photon in the final state.
The analysis uses $t\bar{t}$ events with one of the top quarks decaying to $c+H$ and standard model decays of
the other top quark. The results are corresponding to 19.5 fb$^{-1}$ data at the
center-of-mass energy of 8 TeV. The 95$\%$ C.L. upper limit on $Br (t \rightarrow c H)$ has found to be  5.6$\times 10 ^{-3}$ for a Higgs boson mass of 126 GeV \cite{cmstqh}.
Table \ref{Recent-status} summarizes the current direct limits as well
as the projected ones on the top-Higgs FCNC branching ratios at the
LHC and
the High Luminosity LHC with the center-of-mass energy of 14 TeV and
with the integrated luminosities of
300  and 3000 fb$^{-1}$.
The LHC projections are taken from Ref. \cite{Agashe:2013hma}. As it can be seen
from Table \ref{Recent-status}, the best possible limit on $Br (t \rightarrow q H)$ from the LHC would be at the order of $10^{-4}$ at high luminosity regime.

\begin{table}
	\begin{center}
	\begin{tabular}{c c c c c}
	\hline\hline
	Process & $Br$ Limit & Search & Data set & Reference     \\     \hline
	$t \rightarrow  qH$ & $7.9 \times 10^{-3}$ & ATLAS   $t \rightarrow t \to Wb +  qH \rightarrow  \ell \nu b +\gamma \gamma q$  & 4.7,20 fb$^{-1}$ @ 7,8 TeV &\cite{atlastqh}  \\
	$t \rightarrow  cH$ & $5.1 \times 10^{-3}$ & ATLAS   $t \rightarrow t \to Wb +  qH \rightarrow  \ell \nu b +\gamma \gamma q$  & 4.7,20 fb$^{-1}$ @ 7,8 TeV &\cite{atlastqh}  \\
	$t \rightarrow  cH$ & $5.6 \times 10^{-3}$ & CMS     $t \bar t \rightarrow Wb +  qH \rightarrow  \ell \nu b + \ell \ell q X$   & 19.5 fb$^{-1}$ @ 8 TeV &\cite{cmstqh} \\   \hline
	$t \rightarrow  qH$ & $5 \times 10^{-4}$   & LHC     $t \bar t \rightarrow Wb +  qH \rightarrow  \ell \nu b +\gamma \gamma q$  & 300 fb$^{-1}$ @ 14 TeV   &   \cite{Agashe:2013hma}    \\
	$t \rightarrow  qH$ & $2 \times 10^{-4}$   & LHC     $t \bar t \rightarrow Wb +  qH \rightarrow  \ell \nu b +\gamma \gamma q$  & 3000 fb$^{-1}$ @ 14 TeV  &    \cite{Agashe:2013hma}   \\
	$t \rightarrow  qH$ & $2 \times 10^{-3}$   & LHC     $t \bar t \rightarrow Wb +  qH \rightarrow  \ell \nu b + \ell \ell q X$   & 300 fb$^{-1}$ @ 14 TeV   &    \cite{Agashe:2013hma}  \\
	$t \rightarrow  qH$ & $5 \times 10^{-4}$   & LHC     $t \bar t \rightarrow Wb +  qH \rightarrow  \ell \nu b + \ell \ell q X$   & 3000 fb$^{-1}$ @ 14 TeV  &    \cite{Agashe:2013hma}  \\  \hline \hline
	\end{tabular}
	\end{center}
	\caption{Current direct limits as well as the projected ones on the $Br(t\rightarrow qH)$ at the LHC and future HL-LHC. }	
	\label{Recent-status}
\end{table}

{\it Indirect limits:}
Low energy measurements in flavor mixing processes can be used to
constrain the top quark flavor violation in the $tqH$ vertex.
At loop level, $D^{0}-\bar{D^{0}}$ mixing observable, the mass
difference $\Delta M$,  receives sizeable contributions
from both  $tuH$ and $tcH$ couplings at the same time.  Using the
measured value of $\Delta M$, one can obtain a limit on the product of
two couplings, i.e. $g_{tuH}g_{tcH}$ \cite{dd}. With the Higgs boson
mass in the range of 115-170 GeV,  upper limits of $g_{tuH}g_{tcH}  \leq
(1.94-2.72)\times 10^{-2}$ are obtained. This is corresponding to an upper limit of
$Br(t\rightarrow q H) < (5.3-7.4)\times 10^{-4}$ if we assume $g_{tuH} =
g_{tcH}$.

Another indirect way to constrain the top-Higgs FCNC couplings is
to use the electroweak precision observables of $Z$ bosons
\cite{Larios:2004mx}. The $tcH$ vertex contributes to the $Z\rightarrow
c\bar{c}$ decay at loop level. It affects the electroweak precision
observables in $Zc\bar{c}$ vertex. The total width, partial width, and
asymmetries are affected by the $tcH$ FCNC interaction. In
\cite{Larios:2004mx}, the $tcH$ vertex contribution has been
calculated and the upper limits of $Br(t\rightarrow cH) \leq
(0.09-2.9)\times 10^{-3}$ for the Higgs mass in the range of $114\leq
m_{H} \leq 170$ GeV have been obtained. We update this limit
with the Higgs boson mass of 125 GeV using the current measurements of
$Zc\bar{c}$ vertex.

After taking into account the $tcH$ FCNC coupling contributions to the
width of $Z \rightarrow c \bar{c}$, it can be written as:
\begin{eqnarray}\label{ddu}
\Gamma(Z \rightarrow c \bar{c}) = \Gamma(Z \rightarrow c \bar{c})_{SM}\,(1 + \delta^H_{tcH}) \,,
\end{eqnarray}
where the $tcH$ one loop corrections are given by
$\delta^H_{tcH}$. The details of the calculations of $\delta^H_{tcH}$ are available in \cite{Larios:2004mx}. It can be
expressed in terms of the Veltman-Passarino functions.
Using the calculations and the related inputs from the Particle Data
Group~\cite{Agashe:2014kda}, an upper limit of $Br(t \rightarrow c H)
\leq 2.1 \times 10^{-3}$ is found at $95\%$ C.L.
As it can be seen, the indirect limits are at the same order of the current direct
limits, i.e. $10^{-3}$.

%
%
\section{Study of  $tqH$ in top pair events in e$^{-}$e$^{+}$ collisions}\label{Numerical-study}

\begin{figure}[htb]
\begin{center}
\vspace{1cm}
\resizebox{0.55\textwidth}{!}{\includegraphics{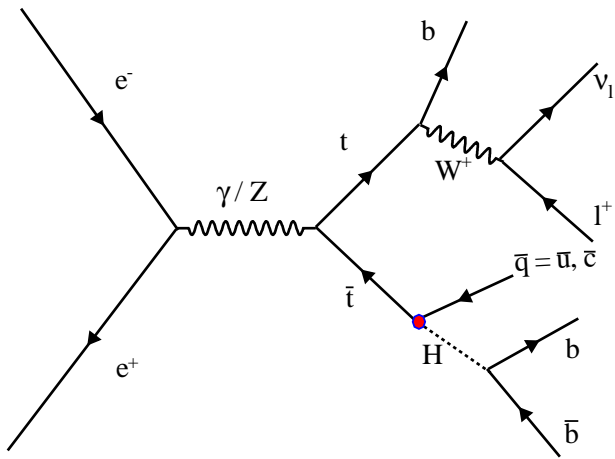}}  
\caption{ The representative Feynman diagram for production of a $t \bar{t}$ event. It includes the decay chain with one top decay leptonically and the other top decay from anomalous FCNC coupling and Higgs decay into a $b \bar{b}$ pair. }\label{fig:feynman}
\end{center}
\end{figure}

In this section, we a search for top-Higgs FCNC couplings in e$^-$e$^+
\rightarrow t (\rightarrow \ell \nu_{\ell} b) \,\, \bar{t}
(\rightarrow q H)$ channel, where $\ell$ = e, $\mu$ and $q=u,c$,
and present the potential of a future electron-positron collider to
probe $tqH$ couplings. As mentioned before, we
concentrate on the semi-leptonic decay of a top quark and
anomalous decay of another top with
the Higgs boson decaying into a $b\bar{b}$ pair, as shown in Fig.\ref{fig:feynman}.
Therefore, the final state consists of an energetic lepton (muon or electron),
neutrino (appears as missing momentum) and four hadronic
jets. Three of the jets are produced from the hadronization of bottom quarks.

In this study, we assume $g_{tqH}^{v} = 1$ and no axial coupling, i.e. $g_{tqH}^{a} = 0$.
At the center-of-mass energy of $\sqrt{s}$ =  500 GeV, the leading order cross section including the branching ratios reads:
\begin{eqnarray}
\sigma_{\sqrt{s} = 500 \, {\rm GeV}}(g_{tqH})~=11.306 \times g_{tqH}^{2} \, \text{(fb)} ;
\label{sigma}
\end{eqnarray}
At higher center-of-mass energy, the cross section decreases as $1/s$.
Now, we turn to event generation and simulation. In order to simulate
the signal events, the top-Higgs FCNC effective Lagrangian
(Eq.\ref{fcnclagrangian}) is implemented in the { \tt FeynRules}
package~\cite{Alloul:2013bka,Christensen:2008py,Duhr:2011se} then the
model is imported to a Universal { \tt FeynRules} Output (UFO)
module \cite{Degrande:2011ua}.  After that, it is inserted to { \tt MadGraph5-aMC@NLO}~\cite{Alwall:2014hca,Alwall:2011uj} event generator.
{ \tt PYTHIA}~\cite{Sjostrand:2007gs,Sjostrand:2003wg} is utilized for
parton showering and hadronization and
{ \tt Delphes 3}~\cite{deFavereau:2013fsa,Mertens:2015kba} is employed
to account for detector effects.

The main background comes from top pair events with
semi-leptonic decay of one of the top quarks and hadronic decay of
another top quark. Other backgrounds to our signal include
$W^{\pm}b\bar{b}jj$, $Zb\bar{b}jj$ (with leptonic decay of $Z$), and
$Z \ell^{\pm}\ell^{\pm}jj$ (with hadronic decay of $Z$). The
contribution of $Wjjjj$ where $j$ denotes non-bottom quark jets is
studied as well. All of these backgrounds are generated at leading order using  { \tt MadGraph5-aMC@NLO}.

To consider detector resolutions, the final state particles, leptons
and jets,  are smeared according to Gaussian distributions using the
following parameterizations which are used in { \tt Delphes 3}.
Jets energies are smeared as ~\cite{Linssen:2012hp,BrauJames:2007aa}:
\begin{eqnarray}\label{energy-smearing-jets}
\frac{\Delta  E_{j}}{E_{j}}  =  \frac{40.0 \%}{\sqrt{E_{j}}} \oplus 2.5\% \,\,\, {\rm (jets)},
\end{eqnarray}
and for leptons (muon and electron), we use a CMS-like detector resolution:
\begin{eqnarray}\label{energy-smearing-lepton}
	\frac{\Delta  E_{\ell}}{E_{\ell}}      =  \frac{7.0\%}{\sqrt{E_{\ell}}}  \oplus \frac{0.35}{E_{\ell}} \oplus 0.7\% \,\,\,  {\rm (leptons)},
\end{eqnarray}
where $E_{j}$ and $E_\ell$ represent the energies of the jets and
leptons, respectively. The energies are in GeV and the symbol $\oplus$
represents a quadrature sum.  It should be mentioned that the
electron and muon energy resolutions are different however, for
simplicity,  we smear the energies of muons similar to the electrons.

The anti-k$_{t}$ algorithm  \cite{kt} with the jet radius of 0.4 is used to
reconstruct jets.
We present the results for three b-jet identification efficiencies of
$\epsilon _{b}$=60$\%,70\%,80\%$. A mistag rate of 10$\%$ for charm
quark jets and 1$\%$ for other light flavor jets are considered.
It is notable that b-tagging efficiency and mistag rates play
important roles in this analysis, as we have b-jets in the final
state as well as light jets.

\begin{figure}[htb]
	\begin{center}
		\vspace{1cm}
		\resizebox{0.70\textwidth}{!}{\includegraphics{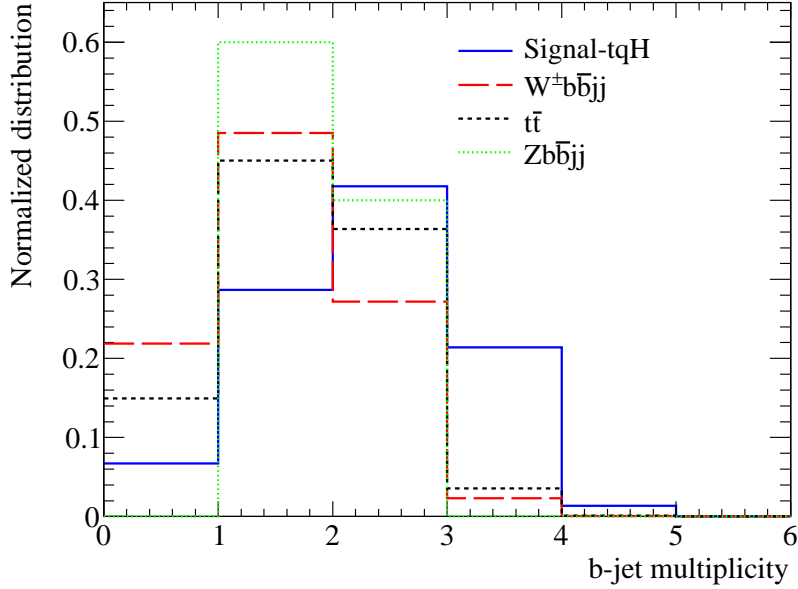}}  
		\caption{Distribution of $b$-jet multiplicity for signal and SM backgrounds.}
		\label{bmultiplisity}
	\end{center}
\end{figure}

The events are selected according to the following strategy.
For each event, to reconstruct the the semi-leptonic decaying top
quark, we require exactly one charged lepton with $p_{T}^{\ell}> 10$ GeV within the pseudo-rapidity range of $\vert\eta^{\ell}\vert < 2.5$.
The events with more than one charged leptons are discarded.
The $W$ boson momentum in the top quark decay is obtained by summing the momenta of the charged lepton and
neutrino.
Each event is required to have exactly four jets, $n_{j} = 4$, with
$p_{T}^{jets} > 20$ and $\vert\eta^{jets}\vert < 2.5$.
Among the jets, at least three jets must be b-tagged jets.
The b-jet multiplicity is presented in Fig.~\ref{bmultiplisity} for signal and
different SM backgrounds. As it can be seen
from the distributions, the three b-tag jets requirement is considerably useful
to reduce contributions of different backgrounds.

To have well isolated objects, for any pair of objects in the final state, we require $\Delta R_{ij}
= \sqrt{(\eta_{i} - \eta_{j})^{2} + (\phi_{i} - \phi_{j})^{2}} > 0.4$
with $i$ and $j$ running over all particles in the final state.

\begin{figure}[htb]
	\begin{center}
		\vspace{1cm}
		\resizebox{0.70\textwidth}{!}{\includegraphics{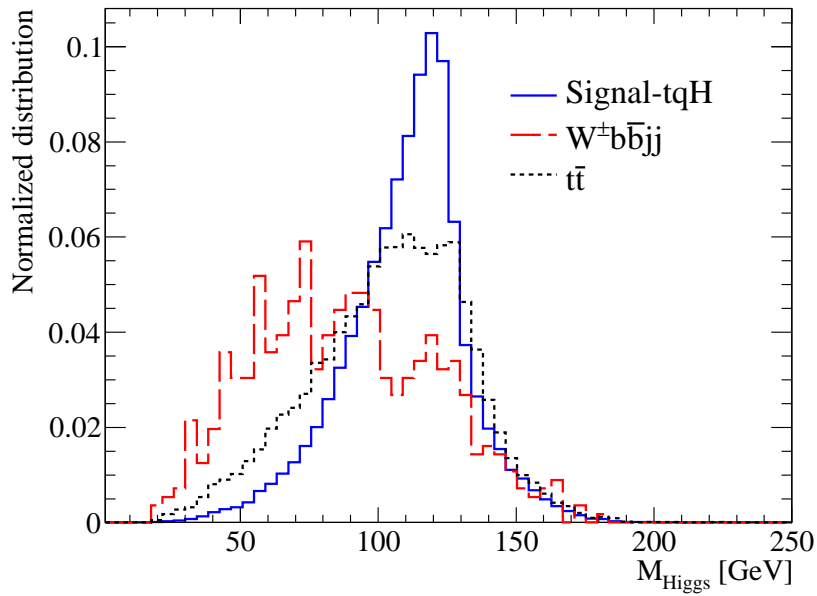}}  
		\caption{The reconstructed Higgs boson mass
                  distribution from the $\chi^{2}$ analysis for signal
                  and bakgorounds. The signal sample is generated with
                  $g_{tuH}$ = 0.5.}
		\label{Higgsmaas}
	\end{center}
\end{figure}

To reconstruct the Higgs boson and then both top quarks, there are
ambiguities to choose the correct combinations of the b-jets.
To solve the ambiguities, and reconstruct the Higgs boson and $t\bar{t}$ system, we define
a $\chi^{2}$ as:
\begin{equation}\label{chi2}
\chi^2_{b_{m} b_{n} b_{k}} = ( m_{b_{m}\, W} - m_{{\rm top}} )^2  + ( m_{b_{n} b_{k}} -  m_{\rm Higgs} )^2,
\end{equation}
Various combinations of $\chi^2_{b_{m}b_{n}b_{k}}$, with $m,n,$
and $k$ run over the b-jets, are made and the one with minimum $\chi^2$ is chosen.
The mass distribution of the reconstructed Higgs boson
is illustrated in Fig.~\ref{Higgsmaas}. As can seen the
invariant mass distribution peaks at the Higgs boson mass for signal
events while backgrounds have wide distributions. As a result,
applying a mass window cut can reduce the
backgrounds contributions. We require the reconstructed invariant mass
of the Higgs boson to satisfy 90 GeV $< m_{\rm Higgs}^{\rm reco} < 140$ GeV.

Table~\ref{Table:Cross-Section} summarizes the cross sections (in fb) after
applying the cuts for the signal and backgrounds. The b-tagging
efficiency is assumed to be $60\%$ in this table.
The contribution of $Zb\bar{b}jj$ (with $Z\rightarrow \ell^{\pm}\ell^{\pm}$) and
$Z \ell^{\pm}\ell^{\pm}jj$ (with $Z\rightarrow jj$) backgrounds is
negiligible after all cuts. After the jets requirements (set II of the
cuts in Table \ref{Table:Cross-Section}), the cross section is at the
order of $10^{-5}$ and goes to zero after the three b-jets requirement.
No events of $W^{\pm}jj jj$, where $j$ denotes light flavor jets,
survives after three b-jet requirement. Therefore, these sources of
backgrounds are not mentioned in Table \ref{Table:Cross-Section}.
Considering different sources of systematic uncertainties in detail is beyond
the scope of this work so an overall uncertainty of $30\%$ is conservatively
assigned to the number of background events for the limit setting procedure.

\begin{table}[htbp]
\begin{center}
\begin{tabular}{c|c|cc}
			$\sqrt s = 500$ GeV         &  \multicolumn{1}{c|}{ Signal }          & \multicolumn{2}{c}{\,\,\,\,\, Backgrounds }    \\   \hline
			 Cuts
                        &  $\sigma_{tqH}$ (fb)
                        & $\sigma_{W^\pm b \bar{b} j j}$ (fb) &  $\sigma_{t
                          \bar{t}}$ (fb)    \\
			\hline  \hline
			No cut                                                &  $11.306\,(g_{tqH})^{2}$             & $1.72$   &  $148.70$       \\
			(I): 1$\ell$ + $|\eta^{\ell}|<2.5$ + $P_T^{\ell}>10$ + $E_T^{miss}>10$      &  $7.972\,(g_{tqH})^{2}$      &     $1.623$  &  $106.065$     \\
			(II): $4 \, jets$ + $|\eta^{jets}|<2.5$ + $P_T^{jets}>20$ + $\Delta R_{\ell,jets} \geq 0.4$  &  $3.399\,(g_{tqH})^{2}$  &  $0.0071$ &  $47.824$        \\
			(III): $n_{b-jet} \geq 3$ + $\Delta R_{\ell,b-jets} \geq 0.4$           &  $0.709\,(g_{tqH})^{2}$        &  $0.00015$   &  $1.417$         \\
			(IV): $90 < m_{\rm Higgs}^{\rm reco} < 140$                                     &  $0.570\,(g_{tqH})^{2}$        &  $0.00005$   &  $0.961$         \\   \hline  \hline
\end{tabular}
\end{center}
\caption{ Cross sections (in fb) after applying different set of cuts for signal and backgrounds. The b-tagging
efficiency is assumed to be $60\%$ in this table. The details of the basic cuts applied, are presented in the text.}
\label{Table:Cross-Section}
\end{table}

Now, we proceed to set the $95\%$ C.L. upper limit on the signal cross
section. Then the limits are translated into the upper limits on $Br(t\rightarrow qH)$.
Upper limits on the signal cross section is calculated with a CL$_{s}$
approach \cite{cls}. The RooStats \cite{roostat} program is utilized
for the numerical evaluations of the CL$_{s}$ limits.

We summarize the $95\%$ C.L. limits on $Br(t\rightarrow qH)$ in Table
\ref{Brresults} for three b-tagging efficiencies of $60\%, 70\%,80\%$ with
300 fb$^{-1}$ and 3000 fb$^{-1}$ of integrated luminosity of data.
With a b-tagging efficiency of $70\%$ and 300 fb$^{-1}$ of data, an upper limit
of $5.894\times 10^{-3}$ could be achieved.
As it can be seen from the Table
\ref{Brresults}, higher b-tagging efficiency leads to improve the
limits at the level of $30\%-40\%$. More amount of data
makes the upper limits better however the gain is less than one order of magnitude.

In comparison with the LHC direct limits
presented in Table \ref{Recent-status}, a
future electron-positron collider would be able to reach
similar sensitivity to the LHC experiments. The limits of the
electron-positron collider could be significantly improved by
including other decay modes of the Higgs boson such as
$H\rightarrow \gamma\gamma$, $W^{+}W^{-}$, and $ZZ$. In addition,
utilizing a more powerful tool, such as a multivariate technique, to separate signal from backgrounds
could provide better the sensitivity.

\begin{table}[htbp]
\begin{center}
\begin{tabular}{cccc}
  \hline
   \head{b-tagging efficiency} & \head{IL} & \head{Upper limit on $g_{tqH}$} & \head{Upper limit on $Br(t\rightarrow qH)$}\\
  \hline    \hline
  $\epsilon_{b} = 60\%$ & 300  fb$^{-1}$ & 0.463 &$5.894 \times 10^{-3}$ \\
  $\epsilon_{b} = 60\%$ & 3000 fb$^{-1}$ & 0.256 &$1.798 \times 10^{-3}$ \\
  $\epsilon_{b} = 70\%$ & 300  fb$^{-1}$ & 0.373 &$3.821 \times 10^{-3}$ \\
  $\epsilon_{b} = 70\%$ & 3000 fb$^{-1}$ &0.202 & $1.126
  \times 10^{-3}$ \\
$\epsilon_{b} = 80\%$ & 300  fb$^{-1}$ &0.301& $2.476 \times 10^{-3}$ \\
$\epsilon_{b} = 80\%$ & 3000 fb$^{-1}$ & 0.166&$7.546 \times 10^{-4}$ \\
  \hline  \hline
\end{tabular}
\end{center}
\caption{ The $95\%$ C.L. limits on $Br(t\rightarrow qH)$ for b-tagging efficiencies of $60\%,70\%,80\%$ with
	300 and 3000 fb$^{-1}$ of integrated luminosity of data.}
\label{Brresults}
\end{table}

%

%
\section{Summary and conclusions}\label{Conclusions}

In this paper, we have presented direct and indirect searches for
top-Higgs FCNC couplings. The radiative corrections due to $tcH$
coupling on the electroweak precision observables of $Z\rightarrow
c\bar{c}$ decay are used to constrain $Br(t\rightarrow cH)$
using the most recent measurements. This leads to the upper limit
of $2.1\times 10^{-3}$ on $Br(t\rightarrow cH)$.

As a direct search, we study a future electron-positron collider
potential at the center-of-mass energy of 500 GeV to search for the
$tqH$ FCNC couplings via top quark pair production.
The search is based on the process in which one of the top quarks
decays leptonically $(t \to b \ell \nu_{\ell})$ and the other top
quark decays anomalously  to $t \to q H$ with Higgs boson decays into
$b \bar{b}$ pairs. The 95$\%$ C.L. upper limits on the branching ratio
of $Br (t \rightarrow q H)$ with $q = u$- and $c$-quark is found to
be   $5.894 \times 10^{-3}$ for 300 fb$^{-1}$ of integrated luminosity
of data. This upper limit decreases down to $1.798 \times 10^{-3}$ for
the 3000 fb$^{-1}$ data. We find that b-tagging efficiency plays an
essential role in this analysis and can improve the results at the
level $30-40\%$ moving from an efficiency of $60\%$ to $70\%$.
These limits could be considerably improved by including the other decay modes of the Higgs boson such as $\gamma\gamma$, $W^{+}W^{-}$, and $ZZ$.

%
%
\section*{Acknowledgments}

The authors thank R. Martinez for the help in updating the limits from $Z$ boson EW precision observables.
Special thank to S. Khatibi for providing the FCNC model for simulating the events in MadGraph.
Authors are thankful School of Particles and Accelerators, Institute for Research in Fundamental Sciences (IPM)
for financially support of this project.
H. Khanpour also thanks the University of Science and Technology of Mazandaran for financial support provided for this research.

%
%

\end{document}